\documentclass[%
 prd, twocolumn,
nofootinbib,
 amsmath,amssymb,
 aps,
]{revtex4-2}

\usepackage{graphicx}% Include figure files
\usepackage{dcolumn}% Align table columns on decimal point
\usepackage{bm}% bold math
\usepackage[colorlinks=true,linkcolor=red,urlcolor=blue,citecolor=blue]{hyperref}
\usepackage{amsmath, amsthm, amssymb, amsfonts,mathrsfs,stackengine,color,slashed,parskip, verbatim,extarrows,needspace,mathtools,bbold,urwchancal,xcolor}

\graphicspath{{figures/}}

\begin{document}

\preprint{APS/123-QED}

\title{On Equation of State of Dark Matter around Massive Black Holes}% Force line breaks with \\

\author{Zhong-Ming Xie$^{a,b,c}$}
\author{Yong Tang$^{c,a,d}$}%
%\email{tangy@ucas.ac.cn}
\affiliation{\begin{footnotesize}
		${}^a$School of Fundamental Physics and Mathematical Sciences, \\
		Hangzhou Institute for Advanced Study, UCAS, Hangzhou 310024, China \\
            ${}^b$Institute of Theoretical Physics, Chinese Academy of Sciences, Beijing 100190, China \\
            ${}^c$University of Chinese Academy of Sciences, Beijing 100049, China \\
            ${}^d$International Center for Theoretical Physics Asia-Pacific, Beijing/Hangzhou, China 
		\end{footnotesize}
        }
\date{\today}

\begin{abstract}
The nature of Dark Matter (DM) remains mysterious despite the substantial evidence from astrophysical and cosmological observations. While the majority of DM in our universe is non-relativistic, collisionless and its equation of state (EoS) is approximately pressureless $p\simeq 0$, DM becomes relativistic near the massive black holes in galactic center. Yet its EoS is seldom discussed in the relativistic regime. Here we initially explore the possible equation of state for DM in the vicinity of Schwarzschild black holes. We work in a spherical and quasi-static background spacetime, and describe DM as a perfect fluid in equilibrium. Through numerically solving the Tolman–Oppenheimer–Volkoff equations with physical boundary conditions, we show that DM can have static profiles near black holes and its pressure should be negative in order to support the viable density profiles $\rho$. We illustrate with two simple general equations of state, namely the power law $p \propto \rho^\gamma$ and the radius-dependent $p \propto r\cdot \rho$, and compare them with the observations of the Milky Way. Our findings provide insights into the model-building of DM, which should incorporate the possibility of negative pressure in the relativistic regime around black holes.
\end{abstract}

\maketitle

\section{Introduction}
Compelling evidence has accumulated over decades, indicating the presence of DM across various cosmic scales~\cite{cirelli2024darkmatter}. Despite these astrophysical and cosmological evidence, the intrinsic nature of DM remains an open question~\cite{Bertone:2004pz}. One approach to exploring the nature of DM is to examine the dark matter halo (DMH) surrounding a massive black hole (BH) in galactic center, as the DMH can affect the emitted gravitational wave (GW) signals and potentially detectable, see~\cite{Barack_2019, Bertone:2019irm, Eda:2013gg, Lacroix:2013qka, Yue:2018vtk, Yue:2019ozq, Kavanagh:2020cfn, Li:2021pxf, Kadota:2023wlm} for examples, and the density profile of the DMH is the main physical quantity to be discussed~\cite{Zhang:2024hrq, Zhao:2024bpp, Cheng:2024mgl, Cao:2024wby, Liang:2022gdk}.

Previous studies had used different methods to discuss DM profile near a central BH. A dense spike could form for the Newtonian and adiabatically grown cold collisionless DM~\cite{PhysRevLett.83.1719}, and its fully relativistic version was developed in~\cite{PhysRevD.88.063522}. A spherically symmetric and static metric functions were solved in~\cite{matos2003generalrelativisticgeometrynavarrofrenkwhite} by the tangential velocity derived from the profile. Similar method was used in a recent work~\cite{Xu_2018}, which considered a linear superposition of the energy-momentum tensors for a Schwarzschild solution and a modeled profile, using the Newman-Penrose method~\cite{Newman:1965tw} to generalize to a rotating BH case. In~\cite{albadawi2024schwarzschildblackholegalaxies}, the energy conditions and time-like geodesics of particles were discussed with a Dehnen-type profile. Also, profiles with attractor behaviors were found in~\cite{10.1111/j.1365-2966.2011.18687.x, 10.1111/j.1365-2966.2011.19258.x} in the Eddington-Finkelstein coordinate. Recently, with a vanishing radial pressure, \cite{PhysRevD.105.L061501} considered a mass profile inspired by the Hernquist profile, and the same method was also used to study the transonic accretion flow in~\cite{patra2025effectdarkmatterhalo}, while several analytical models of profiles are discussed in~\cite{SHEN2024138797, shen2024classanalyticalmodelsblack}.
Earlier, a model of cloud of strings was proposed in~\cite{PhysRevD.20.1294}, and recently in~\cite{hamil2024schwarzschildblackholesurrounded} new development is presented with a perfect fluid DM. In~\cite{PhysRevLett.116.041302, Alvarez:2020fyo} initial values are set far from the center for the Newtonian system to solve the profile. Overall, we shall note that some of these works are mainly based on non fully-relativistic dynamics or a given profile initially.

% Additionally, there are possibilities that the system reaches a thermal quasi-equilibrium, therefore the ODE system may get easier to solve. For instance, the self-interactions between the DM particles may support a system composed with DM to reach such an equilibrium. The work~\cite{PhysRevLett.116.041302} has considered this situation through the hydrostatic equilibrium, where the DM form an isothermal gas near the center.

When gravity gets strong near the center, the relativistic effect is generally not negligible. Therefore, we should solve the coupled Einstein field equations (EFEs). In this sense, the profile is not given initially but solved from the proper boundary conditions (BDC). In this study, we investigate the density distribution of DM surrounding massive black holes (BHs) in galactic center, specifically Sagittarius $A^*$ (Sgr $A^*$) in the Milky Way~\cite{EventHorizonTelescopeCollaboration_2022}. We employ the EFEs with a static, spherically symmetric space-time, and the energy-momentum tensor outside the horizon is sourced by DM in a state of quasi-equilibrium and described by a perfect fluid, which is partially motivated by the self-interacting DM models~\cite{Spergel:1999mh, Tulin:2017ara}. We numerically solve the Tolman–Oppenheimer–Volkoff (TOV) equations~\cite{T, OV}, subjected to the BDC values set far away by the Navarro-Frenk-White (NFW) profile~\cite{NFW}. We use two types of EoS, which are the power law (PL) type $p\propto\rho^\gamma$ and the radius-dependent (RD) type $p\propto r\cdot\rho$. We also assess the system analytically to predict the behavior of the density solutions and show that there exist stable and static DM density profiles outside massive BHs with negative pressure. 

The structure of this paper is as follows. In Sec.~\ref{sec:theo} we establish the theoretical framework and conventions for later discussions. Later in Sec.~\ref{section:num} we introduce two types of EoS, solve TOV equations numerically, and showcase the DM density profiles around a BH. We also discuss the physical criteria that determine the acceptable parameter space. Finally in Sec.~\ref{sec:concl} we give our conclusions.

\section{Theoretical Framework}\label{sec:theo}
%\subsection{The static and spherical space-time metric}
We consider a static and spherical space-time in which the line element is written in a spherical coordinate as
\begin{equation}
    ds^2=-e^{2\Phi(r)}dt^2+e^{2\Lambda(r)}dr^2+r^2d\Omega^2,\label{eq:ds}
\end{equation}
where $\Phi(r)$ and $\Lambda(r)$ are two functions determined by matter distribution. In particular $\Lambda(r)$ is related to the total mass $M(r)$ within radius $r$ through
\begin{equation}
    e^{-2\Lambda(r)}\equiv 1-\frac{2M(r)}{r}.\label{eq:grr}
\end{equation}
Note that in reality spherical symmetry is broken by, for instance the rotation, accretion disk, etc, in which case more complicated spacetime backgrounds are involved. In this paper, we shall first focus on the spherically symmetric case for simplicity. 
%We assume that inside the radius where we take the BDCs, which shall be introduced later, the space-time around the BH is governed by Eq.~\ref{eq:ds}.

We further describe the energy-momentum tensor of DM by a time-independent perfect fluid,
\begin{equation}
    T_{\mu\nu}(r)=\left[\rho(r)+p(r)\right]u_\mu(r) u_\nu(r)+p(r)g_{\mu\nu}(r),
\end{equation}
where $\rho(r)$ is the total mass-energy density of the DM fluid in the local comoving frame, in other words, the sum of the rest mass-energy density and the internal energy density, while $p(r)$ is the pressure in the same frame and $u^\mu(r)$ is the 4-velocity of the fluid element which is normalized, 
\begin{equation}
    u^\mu=e^{-\Phi}(1,0,0,0).
\end{equation}
The continuity condition of the energy-momentum tensor requires
\begin{equation}\label{eq:Phi}
    e^{-2\Lambda}[p^\prime+(p+\rho)\Phi^\prime]=0,
\end{equation}
where the prime ${}^\prime$ denotes the derivative over the radius $r$. With this equation, applying the EFE results in the so-called TOV equations~\cite{T, OV},
\begin{align}
	M^\prime&=4\pi r^2\rho,\label{eq:TOV1}  \\
	p^\prime&=-\frac{M\rho}{r^2}\left(1-\cfrac{2M}{r}\right)^{-1}\left(1+\cfrac{p}{\rho}\right)\left(1+\cfrac{4\pi r^3 p}{M}\right).\label{eq:TOV2}
\end{align}
Here the function $M(r)$ can be regarded as the total energy of the gravitational system. In Eqs.~\ref{eq:TOV1} and \ref{eq:TOV2} there are three unknown quantities, $\rho(r), p(r)$ and $M(r)$. 
%But we only have two independent equations, since Eq.~\ref{eq:Phi} is satisfied automatically through the EFE.
To fully solve the system, we need an additional relation between $p$ and $\rho$, $p=p(\rho,r)$.

After specifying an EoS, $p=p(\rho,r)$, Eq.~\ref{eq:TOV1} and Eq.~\ref{eq:TOV2} become two first-order ordinary differential equations (ODEs) about $M(r)$ and $\rho(r)$. To numerically integrate these two ODEs and get a solution for the mass-energy distribution of the DM, $\rho(r)$, we need to specify two initial values, or BDC. Here we specify the BDC values at $r_{\text{B}}=10^5~r_{\text{BH}}$, where $r_{\text{BH}}\equiv2m_{\text{BH}}$ is the Schwarzschild radii and $m_{\text{BH}}$ is the mass of the central BH,
\begin{align}\label{eq:bdc}
\text{BDCs: }
  \begin{cases} 
    &M_{\text{B}}\equiv M(r_{\text{B}})= m_{\text{BH}} + 1200~{M}_\odot,\\
    &\rho_{\text{B}}\equiv\rho(r_{\text{B}})=\rho_{\text{NFW}}(r_{\text{B}}).
  \end{cases}
\end{align}
We have presumed that $\rho(r)$ at outer region, $r\ge r_{\text{B}}$, is described by the NFW profile~\cite{NFW}, which is expected and supported observationally and numerically for cold DM,
\begin{equation}
    \rho_{\text{NFW}}(r)=\rho_0\left[\dfrac{r}{r_0} \left(1+\dfrac{r}{r_0}\right)^2\right]^{-1}.
\end{equation}
Here $\rho_0$ and $r_0$ are values for density and characteristic length for galaxies, respectively. To be specific, we focus on the central BH of the Milky Way, Sgr $A^*$, taking the parameters as follows. The mass of the central BH $m_{\text{BH}}=4.31\times10^6 ~{M}_\odot$~\cite{centralBHmass}, the characteristic scales in the NFW profile $\rho_0=1.06\times10^{-2}~{M}_\odot/\text{pc}^{3}$ and $r_0=12.5\times10^3~\text{pc}$~\cite{NFWparameters}. Therefore we have $r_{\text{BH}}=4.12\times10^{-7}~\text{pc}$, $r_{\text{B}}=4.12\times10^{-2}~\text{pc}$, and $\rho_{\text{B}}=3.22\times 10^{3}~{M}_\odot/\text{pc}^{3}$.

Also we have used the upper or conservative value $1200~{M}_\odot$ constrained by \texttt{GRAVITY}~\cite{thegravitycollaboration2024improvingconstraintsextendedmass} for the enclosed mass at the radius $\sim 10^{-2}~\text{pc}$ outside the Sgr $A^*$ as $\sim 1200~{M}_\odot\ll m_{\text{BH}}$. The numerical solutions are not sensitive to the BDC mass as we shall discuss in the next section. Therefore we can impose an upper bound,
\begin{equation}
M_{\text{encl}}\equiv\int_{r_{\text{BH}}}^{r_{\text{B}}}dM(r)\lesssim 10^{-3}{M}_\odot \ll m_{\text{BH}}.\label{eq:criterion1}
\end{equation}
Meanwhile, since the time-like and null geodesics only go inwards at the event horizon (EH), we expect the density at the EH is vanishing, 
\begin{equation}\label{eq:criterion2}
    \rho(r_{\text{BH}})=0.
\end{equation}
Eq.~\ref{eq:criterion1} and Eq.~\ref{eq:criterion2} shall be viewed as two criteria for justifying whether a density profile is viable.

\section{Numerical Results}\label{section:num}

\begin{figure}[t]
	\centering
	\begin{minipage}{\linewidth}
		\centering
		\includegraphics[width=\linewidth, height=0.7\textwidth]{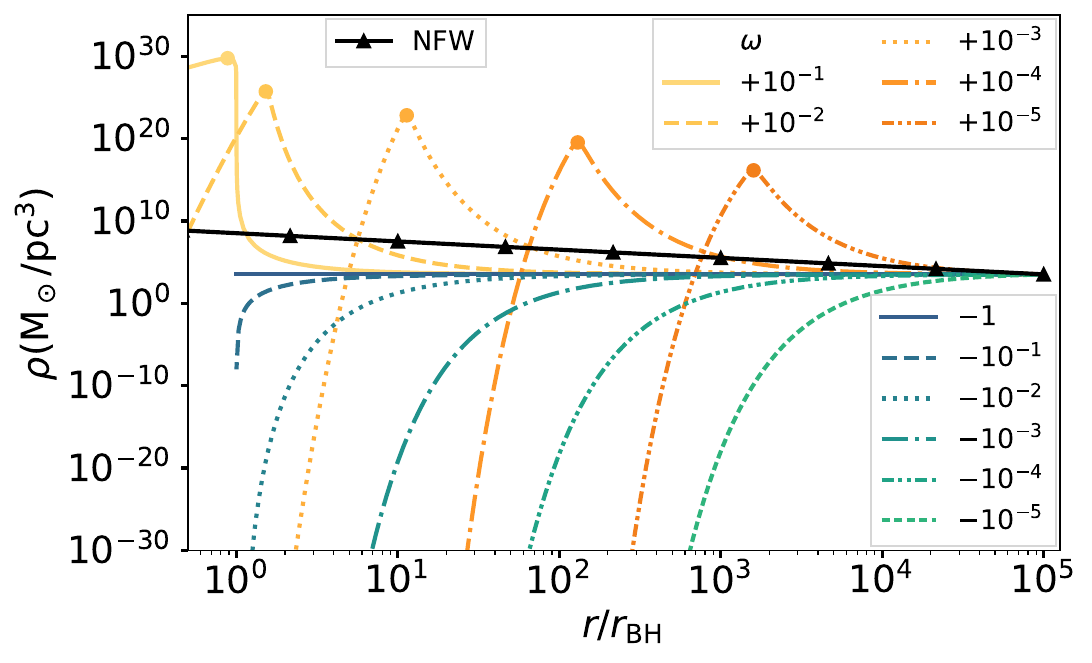}
		\caption{Density profiles for $p=\omega\rho$. Each curve represents a specific value of $\omega$ as indicated. The dots mark the maxima of the density profiles, $\rho_{\text{max}}$ located at $r_{\text{max}}$. Their relations to the mass at the same radius are illustrated in Fig.~\ref{fig:kgammam}.}
		\label{fig:kgammarho}
	\end{minipage}
	
	\begin{minipage}{\linewidth}
		\centering
		\includegraphics[width=\linewidth, height=0.7\textwidth]{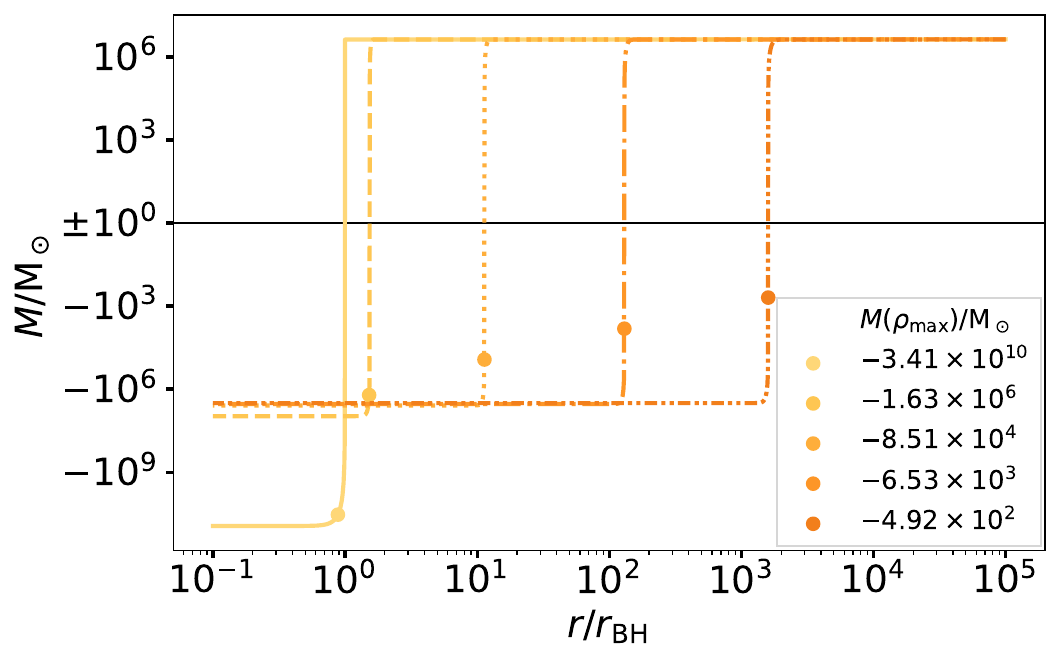}
		\caption{Mass function $M(r)$ for $p=\omega\rho$ with $\omega>0$. Each curve has the same $\omega$ as the one in Fig.~\ref{fig:kgammarho} with the same color; each round dot shares the same radius $r_{\text{max}}$ as the one in Fig.~\ref{fig:kgammarho} with the same color. This figure shows that for each $\rho_{\text{max}}$ located at $r_{\text{max}}$, there is always a negative $M(\rho_{\text{max}})$ at $r_{\text{max}}$.}
		\label{fig:kgammam}

	\end{minipage}
\end{figure}

We consider two different types of EoS in the following discussions. The first one is the PL type which appears in compact stars and cosmology, $p=\omega \rho_{\text{B}}^{1-\gamma} \rho^\gamma$, where $\omega$ is a constant and the factor $\rho_{\text{B}}^{1-\gamma}$ is introduced to make $\omega$ dimensionless. The second EoS is the RD type $p=\zeta\cdot (r/r_{\text{BH}}) \rho$, where $\zeta$ is a dimensionless constant. The motivation is as follows. Generally, the coefficient in front of $\rho^\gamma$ in the PL type EoS is expected to be radius-dependent. In this sense, the next simplest case is a linear dependence with $\gamma=1$.

\subsection{$p=\omega \rho_{\text{B}}^{1-\gamma} \rho^\gamma$}

Now we discuss the PL type EoS. We shall first use the BDC value as $M_{\text{B}}=m_{\text{BH}}+1200{M}_\odot $, and then discuss about how different BDC values $M_{\text{B}}\simeq m_{\text{BH}}$ affect the solutions, relate to the sensitivity of the ODE system.
%And we are going to discuss about the cases with $\gamma=1$ first, then the general cases with $\gamma>0$ later.

$\gamma=1$: We have the EoS $p=\omega \rho$ where $\omega$ can be negative and positive. Due to the weak energy condition (WEC), $p/\rho>-1$, we only consider the cases with $\omega>-1$. We show the corresponding density profiles and the mass functions for different $\omega$ in Fig.~\ref{fig:kgammarho} and Fig.~\ref{fig:kgammam}, respectively~\footnote{Although the figures extend the region inside the EH, the behaviors of the curves inside the EH is not physically reliable.}. Also, we have checked that the curves solved from different BDC radius $r_{\text{B}}=10^3 r_{\text{BH}}$ and $r_{\text{B}}=10^4 r_{\text{BH}}$ have the similar behaviors as shown in these two figures.

As shown in Fig.~\ref{fig:kgammam} for the cases with $\omega>0$, those solutions satisfying Eq.~\ref{eq:criterion2} would surprisingly lead to a negative total mass, $M(r)<0$, for some $r\in(r_{\text{BH}},r_{\text{B}})$. This may be understood as follows. The negative mass indicates that, if we try to solve the system by taking two mass BDC values as $M(r_{\text{B}})=m_{\text{BH}}+1200 {M}_\odot $ and $M(r_\text{BH})=m_{\text{BH}}$, instead of taking one mass BDC value plus one density BDC value in Eq.~\ref{eq:bdc}, no solution consistent with Eq.~\ref{eq:criterion1} could be found for $\omega>0$. This can be seen more transparently in Fig.~\ref{fig:kgammam} where large negative mass is present in inner region to compensate huge positive mass outside and get a positive $M_{\text{B}}$. Then these density curves for $\omega>0$ plotted in Fig.~\ref{fig:kgammarho} are not physically viable. This claim is mainly based on the behavior of the curves or numerical solutions at this stage. Later we shall conduct a general analysis for the TOV equations in this section for $\gamma >0$, and give a similar result there.

Meanwhile, density solutions for negative pressure with $\omega\in(-1,0)$ are also displayed in Fig.~\ref{fig:kgammarho}. The physical picture of such profiles is that the attractive force exerted on DM from BH and inner DM is balanced by that from outer layer DM due to negative pressure. We see that for $\omega=-1$, the solution for $\rho$ is a homogenous one. And for $\omega=0$, there is no matter distribution, $\rho(r)=0$. In between, we expect that the density curve $\rho$ changes continuously over $\omega$ from $-1$ to $0$. This is confirmed by the behaviors seen from the blue curves to green ones which is shallower than the NFW profile. 

$\gamma >0$: We shall generalize the above discussions to the cases with $\gamma>0$. Firstly, we write Eq.~\ref{eq:TOV2} as
\begin{equation}\label{eq:omegatilderho}
    \rho^\prime=\frac{\rho\cdot(\rho+p)}{\gamma |p|\cdot r^2\left(1-\frac{2M}{r}\right)}
     \cdot\left\{-\mathsf{sgn}[\omega]\cdot\mathscr E(r)\right\},
\end{equation}
where $\mathsf{sgn}$ is the sign function and $\mathscr E(r)\equiv M(r)+4\pi r^3 p(r)$ is defined for convenience. We only consider WEC is preserved, $p/\rho>-1$. Then all the quantities outside the curly braces in Eq.~\ref{eq:omegatilderho} are positive, therefore for all $ r\in (r_{\text{BH}},r_{\text{B}})$,
\begin{equation}\label{eq:sgn_rhopr}
    \mathsf{sgn}[\rho^\prime(r)]=-\mathsf{sgn}[\omega]\cdot\mathsf{sgn}[\mathscr E(r)].
\end{equation}
Note that this relation breaks if WEC fails to hold.
%While the validness about WEC shall be checked later, but until then we will assume it holds. Next we discuss the cases for positive $\omega$ and negative $\omega$ separately.

For $\omega>0$, WEC always holds, therefore $\rho^\prime$ has the opposite sign of $\mathscr E$. On the boundary where $M_{\text{B}}\in(m_{\text{BH}},m_{\text{BH}}+1200{M}_\odot )$, we have
\begin{equation}\label{eq:mathscrE_at_boundary}
    \begin{aligned}
    \mathscr E(r_{\text{B}})
    &=M_{\text{B}}+10^{15} \omega\cdot 4 \pi  r_{\text{BH}}^3 \rho_{\text{B}}
    % &\simeq m_{\text{BH}}\left[1+10^{9}\omega\cdot \frac{4 \pi  r_{\text{BH}}^3 \rho(r_{\text{B}})}{m_{\text{BH}}}\right]\\
    \simeq m_{\text{BH}}\left(1+10^{-7} \omega\right),
\end{aligned}
\end{equation}
where we have used the quantity
\begin{equation}\label{eq:BDCestimation}
    4\pi r_{\text{BH}}^3\rho_{\text{B}}/m_{\text{BH}}=6.59\times 10^{-22}.
\end{equation}
$\mathscr E(r_{\text{B}})>0$, therefore $\rho^\prime(r_{\text{B}})<0$. If $\mathscr E>0$ everywhere outside the EH, then the value of $\rho$ is larger near the EH. However this would be inconsistent with the criterion Eq.~\ref{eq:criterion2}, unless $\mathscr E$ turns negative somewhere outside the EH and keeps being negative as the radius gets smaller. Since $4\pi r^3 p>0$, this requires the existence of a negative value of $M$ somewhere outside the EH to compensate the positive value of $4\pi r^3 p$ and make $\mathscr E<0$. As commented above, these solutions with negative total mass-energy are not viable physically, therefore the cases for $\omega>0$ are disfavored.

For $\omega<0$, we only consider $\omega\in(-1,0)$, since $\omega$ in this region do not break WEC at $r_{\text{B}}$. We shall now work with WEC being satisfied inside the boundary and check its validness later. Now $\rho^\prime$ has the same sign with $\mathscr E$. Also we have $\mathscr E(r_{\text{B}})>0$ as calculated in Eq.~\ref{eq:mathscrE_at_boundary}, therefore $\rho^\prime(r_{\text{B}})>0$. With a positive $M$ everywhere outside the EH, $\rho^\prime$ would stay positive as the radius gets smaller and reach to $0$ at $r_{\text{BH}}$ because $\rho(r_{\text{BH}})=0$, as promised by the factor $1-2M/r$ in Eq.~\ref{eq:omegatilderho}. This agrees with our criterion, unless $\mathscr E$ switches its sign somewhere outside the EH. To exclude this possibility, we assume that outside the EH there exists $r_1$, such that $\mathscr E(r_1)=0$, this gives us $(\rho(r_1)/\rho_{\text{B}})^\gamma\simeq(m_{\text{BH}}/4\pi r_{\text{BH}}^3\rho_{\text{B}})/(-\omega\cdot  r^3/r_{\text{BH}}^3)\gg1$, according to Eq.~\ref{eq:BDCestimation}. This equality would never hold for $\gamma>0$, since $\rho^\prime>0$ in the region $(r_1,r_{\text{B}})$ and $\rho(r_1)/\rho_{\text{B}}<1$. Therefore $\mathscr E>0$ and $\rho^\prime>0$ everywhere outside the EH for $\omega\in(-1,0)$. 

Now we have to check the validness of WEC inside the boundary. For $\gamma\geq1$, as the density decreases when the radius is getting smaller, $p/\rho=\omega(\rho/\rho_{\text{B}})^{\gamma-1}$ gets greater, therefore $p/\rho>-1$ always holds. In this case our discussion above is self-consistent.

\begin{figure}[t]
	\centering
	\begin{minipage}{\linewidth}
		\centering
		\includegraphics[width=\linewidth, height=0.7\textwidth]{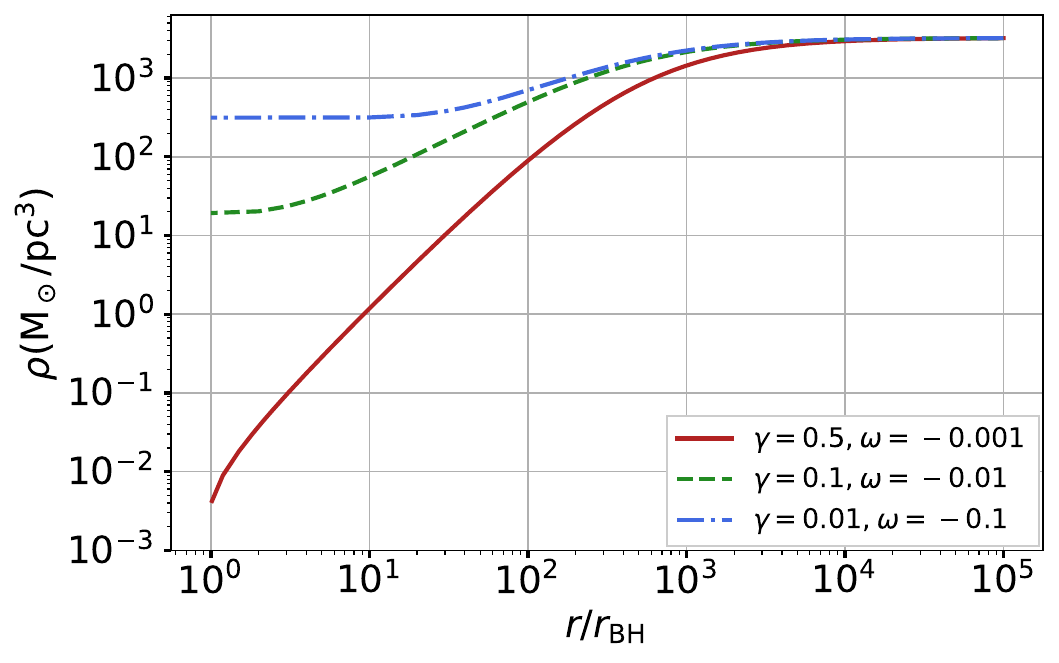}
        \caption{Verification of Eq.~\ref{eq:rho_rc_relation}. The corresponding values of $\rho(r_{\text{BH}})$ in units of ${M}_\odot/\text{pc}^{3}$ are calculated to be $3.22\times 10^{-3}$ for $\gamma=0.5$ and $\omega=-0.001$, $1.93\times 10^{1}$ for $\gamma=0.1$ and $\omega=-0.01$, and $3.15\times 10^{2}$ for $\gamma=0.01$ and $\omega=-0.1$. These values are in agreement with the numerical results presented in this graph, thereby validating Eq.~\ref{eq:rho_rc_relation}.}
		\label{fig:gamma0to1check}
	\end{minipage}
	
\end{figure}

For $0<\gamma<1$, $p/\rho=\omega(\rho_{\text{B}}/\rho)^{1-\gamma}$ gets smaller as the radius gets smaller. Therefore WEC holds as the radius gets smaller starting from the boundary, until the radius reaches $r_{\text{c}}$ where $p(r_{\text{c}})/\rho(r_{\text{c}})=-1$. The existence of $r_{\text{c}}>r_{\text{BH}}$ is ensured by the term $1-2M/r$ in Eq.~\ref{eq:omegatilderho}, which causes $\rho^\prime$ to become sufficiently large as one approaches the EH. At $r_{\text{c}}$, both $\rho^\prime$ and $p^\prime= \rho^\prime\cdot \gamma p/\rho$ vanish due to Eq.~\ref{eq:omegatilderho}. Moreover, all higher-order derivatives of $\rho$ and $p$ also vanish at this point. This is because they are composed of terms that are either directly proportional to $(\rho + p)$ or proportional to the lower-order derivatives of $(\rho + p)$, which are all zero at $r_{\text{c}}$. Therefore inside $r_{\text{c}}$, the density is a constant. Meanwhile, we have $\rho(r_{\text{c}})/\rho_{\text{B}}=|\omega|^{1/(1-\gamma)}$ and $\rho(r_{\text{BH}})=\rho(r_{\text{c}})$, therefore
\begin{equation}\label{eq:rho_rc_relation}
    \rho(r_{\text{BH}})=|\omega|^{\frac{1}{(1-\gamma)}}\rho_{\text{B}}.
\end{equation}
This relation is confirmed in Fig.~\ref{fig:gamma0to1check} for some specific values of $\gamma$. This nonzero $\rho(r_{\text{BH}})$ is against our criterion Eq.~\ref{eq:criterion2}, therefore the cases with $0<\gamma<1$ and $\omega\in(-1,0)$ are not viable.

\begin{figure}[t]
	\centering
	\begin{minipage}{\linewidth}
		\centering
		\includegraphics[width=\linewidth, height=0.7\textwidth]{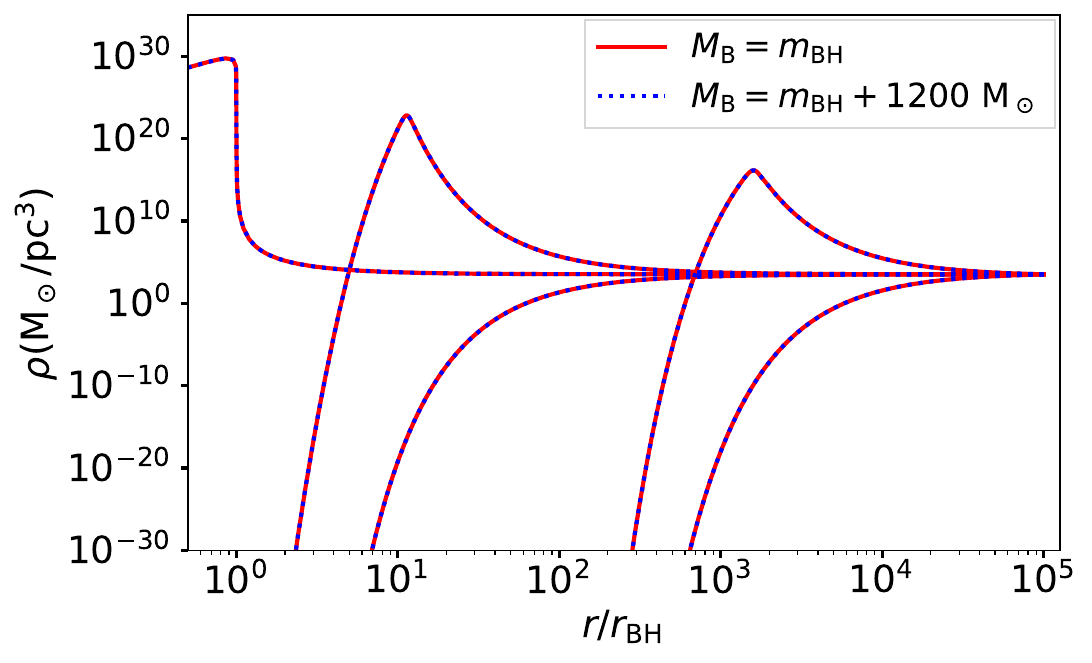}
		\caption{Comparison between the profiles with $M_{\text{B}}=m_{\text{BH}}$ (solid lines) and $M_{\text{B}}=m_{\text{BH}}+1200{M}_\odot $ (dotted lines), two nearly overlapped curves share the same $\omega$, where the EoS is $p=\omega\rho$.}
		\label{fig:mBDCcompare}
	\end{minipage}
\end{figure}

Now we discuss about the choice of the BDC mass value $M_\text{{B}}\in(m_\text{{BH}},m_\text{{BH}}+1200{M}_\odot)$. Our general conclusion for $\gamma>0$ drawn above is based on the form of Eq.~\ref{eq:TOV2} and the sign of $\rho^\prime$ at $r_{\text{B}}$, while the latter is related to $M_\text{{B}}$ through the sign of the quantity $\mathscr E$. Since $\mathscr E(r_{\text{B}})>0$ for all $M_\text{{B}}$ according to Eq.~\ref{eq:mathscrE_at_boundary} when WEC is satisfied, therefore we can always draw the same general conclusion for different choices of $M_\text{{B}}$. Also, as Fig.~\ref{fig:mBDCcompare} shows, the numerical density solutions is not sensitive to $M_\text{{B}}$ for the EoS $p=\omega\rho$.

TOV equations indicate that there is always a homogeneous solution for $\rho$ when $\omega=-1$ since then $\rho_\text{B}+p(r_\text{B})=0$. And there is always a solution with no matter distribution when $\omega=0$. We expect that the curve $\rho$ to smoothly change as $\omega$ varies for a given $\gamma$.

As a conclusion for the cases with $\gamma>0$, scenarios with $p>0$ are not viable, as they at least violate one criterion of Eq.~\ref{eq:criterion1} and Eq.~\ref{eq:criterion2}. And for $p<0$, to satisfy the WEC we only consider $\omega\in(-1,0)$, where we find acceptable solutions with $\gamma\geq1$, while cases with $0<\gamma<1$ break our criterion Eq.~\ref{eq:criterion2} and are therefore not viable. The acceptable solutions all exhibit the same behaviors that density decreases as it approaches the EH. Also, this conclusion is valid for all $M_\text{{B}}\in(m_\text{{BH}},m_\text{{BH}}+1200{M}_\odot)$, and the ODE sensitivity for the EoS $p=\omega\rho$ is low as Fig.~\ref{fig:mBDCcompare} shows. At last, we can check that the enclosed mass, 
\begin{equation}\label{eq:enclmass_esimation_kgamma}
    M_{\text{encl}}<4\pi r_{\text{B}}^3 \rho_{\text{B}}/3\simeq 10^{-1}{M}_\odot,
\end{equation}
is consistent with the upper limit set in~\cite{thegravitycollaboration2024improvingconstraintsextendedmass}.

\subsection{$p=\zeta\cdot (r/r_{\text{BH}}) \rho$}
Next we discuss the RD type EoS. Eq.~\ref{eq:TOV2} is now written as
\begin{equation}\label{eq:rhopr_compact}
    \rho^\prime=-\frac{\rho}{\zeta r \left(1-\cfrac{2M}{r}\right)}\cdot (\Pi_1+\Pi_2), 
\end{equation}
where for later convenience, we have defined two quantities as 
\begin{subequations}\label{eq:Pi1_Pi2}
    \begin{align} 
    \Pi_1 &\equiv  1-\left(1-\frac{r_{\text{BH}}}{\zeta\cdot r}\right)\frac{M}{r}\equiv 1-\left(1-\frac{1}{\zeta\cdot \bar r}\right)\frac{\bar{M}}{2 \bar r},\\
    \Pi_2 &\equiv  4\pi r^2\rho \left(1+\frac{\zeta\cdot r}{r_{\text{BH}}}\right)\equiv \frac{\beta}{2}\bar r^2 \bar\rho(1+\zeta\cdot \bar r).
    \end{align}
\end{subequations}
Where we have defined the dimensionless quantities $\bar{M}\equiv M/m_{\text{BH}}$, $\bar r\equiv r/r_{\text{BH}}$, and $\bar \rho\equiv \rho/\rho_\star, \beta\equiv 4\pi r_{\text{BH}}^3 \rho_\star/m_{\text{BH}}$. Here $\rho_\star>0$ is a constant pivot density and its value shall be chosen later. Eq.~\ref{eq:rhopr_compact} shows that the sign of $\rho^\prime$ is determined only through the signs of $\zeta$ and $\Pi_1+\Pi_2$, since the terms $\rho$ and $r(1-2M/r)$ are positive.

In the cases of $\zeta>0$, $\rho^\prime$ has an opposite sign with $\Pi_1+\Pi_2$. With $M>0$ everywhere outside the EH, we have $\Pi_1>1-2M/r>0$. Since $\Pi_2>0$, therefore $\rho^\prime<0$. These cases with $\zeta>0$ are therefore not viable for the same reason as discussed before in the PL type EoS.

Next, we consider $\zeta<0$. Now $\rho^\prime$ has the same sign as $\Pi_1+\Pi_2$. Since $\bar r\in (1,10^5)$, we may only consider $\zeta\in(-10^{-5},0)$ to satisfy the WEC, namely $p/\rho=\zeta\cdot \bar r\in(-1,0)$. We see that $\Pi_1\in(-\infty,1-2M/r)$ for $M>0$ everywhere outside the EH and $\Pi_2>0$. Therefore a positive $\Pi_1$ indicates a density solution that decreases towards smaller radius, while a negative $\Pi_1$ may result in an extremum of $\rho$. This extremum can be a global maximum due to the negativity of the boundary value $\rho^\prime(r_\text{B})$. According to the numerical results we present later, this is the situation that we shall focus on.

\begin{figure}[t]
    \begin{minipage}{\linewidth}
        \centering
        \includegraphics[width=\linewidth, height=0.7\textwidth]{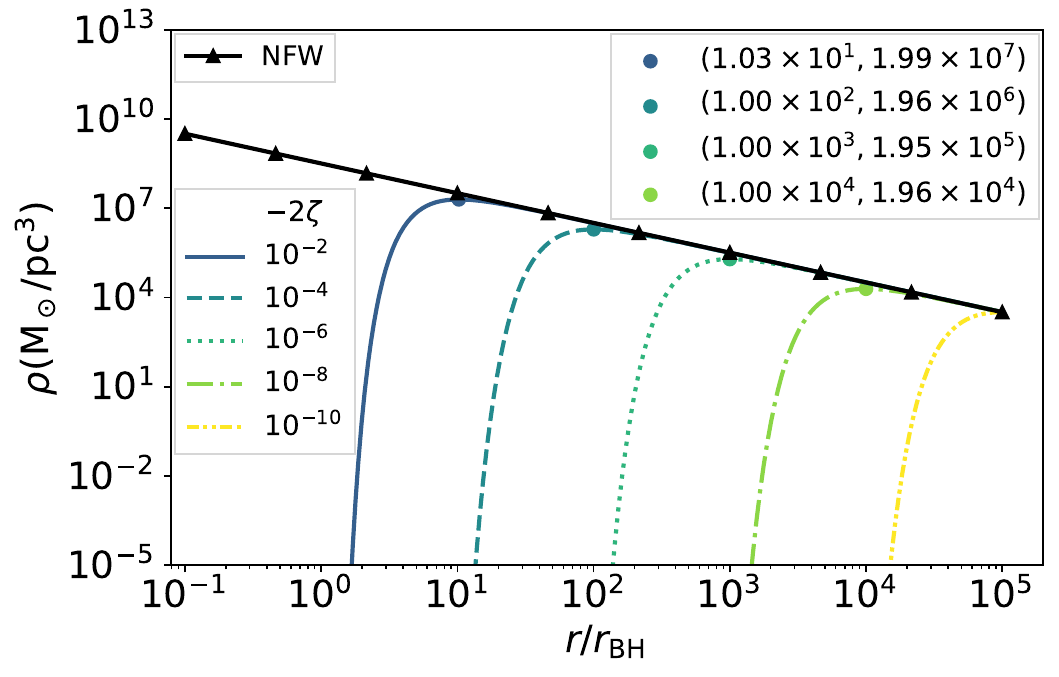}
        \caption{Density profiles for $p=\zeta\cdot (r/r_{\text{BH}}) \rho$. The legend for the round dots represents the coordinates of these maxima, $(\bar r_\text{m},\rho(\bar r_{\text{m}}))$.}
        \label{fig:zetaminu1over40rho}

	\end{minipage}
    \begin{minipage}{\linewidth}
		\centering
		\includegraphics[width=\linewidth, height=0.7\textwidth]{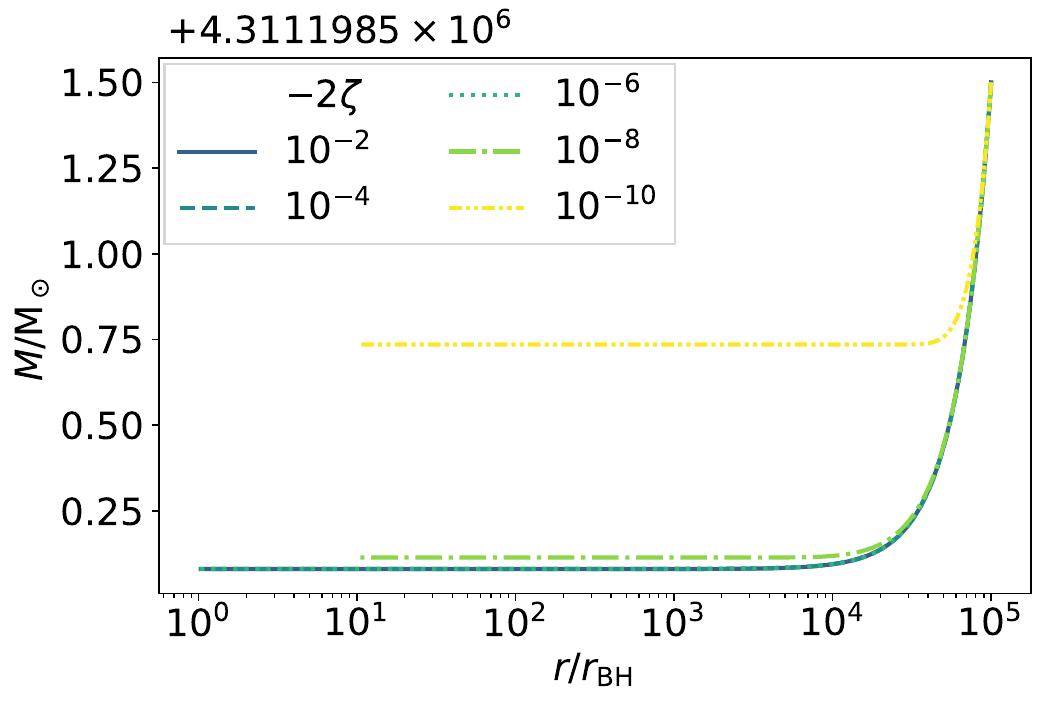}
		\caption{Mass functions for $\zeta$ in Fig.~\ref{fig:zetaminu1over40rho}. The enclosed mass is roughly as $1~{M}_\odot \ll 10^3~{M}_\odot $. Note that the vertical axis has indicated the deviation from the base value $M_{\text{B}}-1.5 ~M_\odot=(4.31\times10^6+1198.5)~M_\odot$.}
        \label{fig:zetaminu1over40M}
	\end{minipage}
\end{figure}

When the maximum exists, we can estimate the corresponding $\zeta$. We first denote the radius of the maximum by $\bar r_{\text{m}}$, and the corresponding values $\bar \rho_{\text{m}}\equiv \bar \rho(\bar r_{\text{m}})$ and $\bar{M}_{\text{m}}\equiv \bar{M}(\bar r_{\text{m}})$ for convenience. Then at $\bar r_{\text{m}}$, the density gradient vanishes, equivalently $\Pi_1(\bar r_{\text{m}})+\Pi_2(\bar r_{\text{m}})=0$,
\begin{equation}
    \alpha \bar r_{\text{m}}^2 \zeta^2+\left(\frac{2\bar r_{\text{m}}}{\bar{M}_{\text{m}}}-1+ \alpha\right)\bar r_{\text{m}} \zeta+1=0. \label{eq:zeta_quadratic}
\end{equation}
Here we have defined $\alpha\equiv \beta\bar r_{\text{m}}^3\bar \rho_{\text{m}}/\bar{M}_{\text{m}}>0$. 

Since $\alpha \bar r_{\text{m}}^2>0$, this quadratic equation of $\zeta$ has a negative solution if and only if $(2\bar r_{\text{m}}/\bar{M}_{\text{m}}-1+ \alpha)\bar r_{\text{m}}>0$ and $[(2\bar r_{\text{m}}/\bar{M}_{\text{m}}-1+ \alpha)\bar r_{\text{m}}]^2-4\cdot\alpha \bar r_{\text{m}}^2\geq 0$. These further give two conditions,
\begin{subequations}
    \begin{gather}  
	\frac{2 \bar r_{\text{m}}}{\bar{M}_{\text{m}}}>1-\alpha,\label{eq:b_g_0} \\
	\left(\frac{2\bar r_{\text{m}}}{\bar{M}_{\text{m}}}-1+\alpha\right)^2\geq 4\alpha.\label{eq:Delta_geq_0}
    \end{gather}
\end{subequations}
Since $2\bar r_{\text{m}}/\bar{M}_{\text{m}}>2/\bar{M}_{\text{m}}>1>1-\alpha$, Eq.~\ref{eq:b_g_0} is satisfied automatically. So would Eq.~\ref{eq:Delta_geq_0}, if we only consider the situation when $\bar r_{\text{m}}\geq \bar{M}_{\text{m}}= 1+\mathcal{O} ( 10^{-3} )$, which is the case we are interested in.

We can now solve Eq.~\ref{eq:zeta_quadratic} for $\zeta$,
\begin{equation}\label{eq:solution_zeta}
    \zeta=\frac{\left(\dfrac{2\bar r_{\text{m}}}{\bar{M}_{\text{m}}}-1+ \alpha\right)\pm \sqrt{\left(\dfrac{2\bar r_{\text{m}}}{\bar{M}_{\text{m}}}-1+ \alpha\right)^2-4\alpha }}{-2 \alpha \bar r_{\text{m}}}.
\end{equation}
Given that $\alpha\ll 1$, which is confirmed later, then Eq.~\ref{eq:solution_zeta} can be expanded. According to the smallness of $|\zeta|\in(0,10^{-5})$, we only take the root with the minus sign,
\begin{align}
        \zeta
        \simeq-\frac{1}{\bar r_{\text{m}}\left(2\bar r_{\text{m}}/{\bar{M}_{\text{m}}}-1\right)}
        \simeq-\frac{1}{\bar r_{\text{m}}\left(2\bar r_{\text{m}}-1\right)},\label{eq:zeta_barr}
\end{align}
where $\bar{M}_{\text{m}}=1+\mathcal{O} ( 10^{-3} )$ is used and all higher-order terms are dropped. 

To make sense of this solution, $\alpha\ll 1$ must hold. We choose $\rho_\star=\rho_{\text{B}}$, therefore $\beta\sim 10^{-22}$ according to Eq.~\ref{eq:BDCestimation}. $\alpha\ll 1$ is now equivalent to
\begin{equation}\label{eq:rhom_constaint}
    \begin{aligned}
        \rho(\bar r_{\text{m}})&\ll \rho_{\text{B}}/(\bar r_{\text{m}}^{3}\beta)
        \simeq 10^{25}\bar r_{\text{m}}^{-3}\cdot {M}_\odot/ \text{pc}^{3}.
    \end{aligned}
\end{equation}

Furthermore, as showed in numerical solutions, $\rho^\prime(r_{\text{B}})<0$ is necessary for the maximum to occur. This condition is equivalent to solving a quadratic equation of $\zeta$, whose solution reads $-2\times10^{7}\lesssim\zeta\lesssim-10^{-10}/2$. Together with the WEC, the condition for a maximum gives
\begin{equation}
    -10^{-5}<\zeta\lesssim-10^{-10}/2.\label{eq:zeta_bounds}
\end{equation}

Using Eq.~\ref{eq:zeta_barr}, we can get the values of $\zeta$ corresponding to different locations of the maxima $\bar r_\text{m}$. 

The above analysis is confirmed by the numerical results in Fig.~\ref{fig:zetaminu1over40rho}. It shows that, the density profile has no maximum for $\zeta\gtrsim-10^{-10}/2$. And for $-10^{-5}<\zeta\lesssim-10^{-10}/2$ the profile has a maximum and satisfies the WEC, while for $\zeta\leq-10^{-5}$ having a maximum but violating the WEC. It also shows that the density profiles are continuously changing over $\zeta$ within its bounds and shallower than the NFW profile. Meanwhile, the smallness of $\alpha$ or Eq.~\ref{eq:rhom_constaint} equivalently, can also be checked and confirmed. At last, in Fig.~\ref{fig:zetaminu1over40M} the corresponding mass functions confirm the enclosed masses are indeed small but positive. 

\section{conclusion}\label{sec:concl}
We have investigated the possible static density profiles of DM near massive black holes in galactic center. To be specific, we have illustrated with Sgr $A^*$ in the Milky Way center. Treating DM as a perfect fluid motivated by self-interacting DM models, and solving the relativistic TOV equations, we have shown that negative pressure near black holes is crucial for DM to form static profiles, which are typically shallower than the NFW profile.

\begin{comment}
This study presents a comprehensive analysis on the possible EoS of DM, focusing specifically on a system which is assumed to be predominantly composed of DM and is near the massive BH in Milky Way center, Sgr $A^*$. Although baryonic matter is the dominant species in reality, our treatment provides valuable insights. Under the assumption of a static and spherically symmetric space-time, and considering the DM as a perfect fluid without fluid flow, we solve the EFEs and formulate the TOV equations~\cite{T, OV}. The BDCs are imposed far from the center, such that the density adheres to the NFW profile~\cite{NFW} and the mass is closed to the upper limit~\cite{thegravitycollaboration2024improvingconstraintsextendedmass}. Along with these set-ups, two criteria are proposed to determine whether a solution is viable.    
\end{comment}

We have discussed two types of EoS, $p\propto\rho^\gamma$ and $p\propto r\cdot\rho$, and visualized the corresponding density and mass profiles. Also we have analyzed the TOV equations analytically with the weak energy condition and confirmed with numerical results. Our findings are based on a fully relativistic treatment, and the profiles are solved from the BDC given far from the center, not given by hand initially. To probe the DM profiles near massive black holes, future gravitational-wave detectors might be useful as the extreme-mass-ratio inspirals would be able to probe the environmental effects~\cite{Li:2021pxf, LISA:2022yao}. 

\section*{acknowledgement}
This work is supported by the National Key Research and Development Program of China (Grant No.2021YFC2201901) and the Fundamental Research Funds for the Central Universities. 

\bibliography{ref}
\end{document}